# Protein identification with deep learning: from abc to xyz

Ngoc Hieu Tran[1], Zachariah Levine[1], Lei Xin[2], Baozhen Shan[2], Ming Li[*, 1]

[1] David R. Cheriton School of Computer Science, University of Waterloo, Waterloo, Ontario, Canada.

[2] Bioinformatics Solutions Inc., Waterloo, Ontario, Canada.

* Corresponding author.

Email: mli@uwaterloo.ca

# ABSTRACT

Proteins are the main workhorses of biological functions in a cell, a tissue, or an organism. Identification and quantification of proteins in a given sample, e.g. a cell type under normal/disease conditions, are fundamental tasks for the understanding of human health and disease. In this paper, we present DeepNovo, a deep learning-based tool to address the problem of protein identification from tandem mass spectrometry data. The idea was first proposed in the context of *de novo* peptide sequencing [1] in which convolutional neural networks and recurrent neural networks were applied to predict the amino acid sequence of a peptide from its spectrum, a similar task to generating a caption from an image. We further develop DeepNovo to perform sequence database search, the main technique for peptide identification that greatly benefits from numerous existing protein databases. We combine two modules *de novo* sequencing and database search into a single deep learning framework for peptide identification, and integrate de Bruijn graph assembly technique to offer a complete solution to reconstruct protein sequences from tandem mass spectrometry data. This paper describes a comprehensive protocol of DeepNovo for protein identification, including training neural network models, dynamic programming search, database querying, estimation of false discovery rate, and de Bruijn graph assembly. Training and testing data, model implementations, and comprehensive tutorials in form of IPython notebooks are available in our GitHub repository (https://github.com/nh2tran/DeepNovo).

# INTRODUCTION

While most human cells carry the same genome regardless of their types or developmental stages, the proteome, the complete set of proteins in a given cell, is much more dynamic due to differential gene expression, alternative splicing, or post-translational modifications. Proteomics is the large-scale study of the proteome expressed in a given biological sample. Liquid chromatography coupled with tandem mass spectrometry (LC-MS/MS) is the current state-of-the-art technology for protein characterization in proteomics [2-3]. In those experiments, proteins are first digested into peptides by different enzymes (simply speaking, a long amino acid sequence is cleaved into several short subsequences), and the peptides are subsequently analyzed by LC-MS/MS instruments. Typical analysis workflows of LC-MS/MS data, e.g. PEAKS, OpenMS, include peptide feature detection and quantification from an LC-MS map, peptide identification from MS/MS spectra, and protein profiling [4-5].

In this paper, we focus on the problem of protein identification that includes (i) predicting peptide sequences from MS/MS spectra and (ii) assembling those peptides back to original protein sequences. The task of peptide identification is to predict the amino acid sequence of a peptide given its spectrum and its mass (Figure 1). There are 20 standard amino acids and their names are often abbreviated by single alphabet letters. Each amino acid has its own molecule mass and the peptide mass is the total mass of its amino acids plus its N-terminal and C-terminal. A spectrum is a collection of (mass, intensity) of ions resulted from the fragmentation at the peptide backbone between any two adjacent amino acids. For example, consider the peptide "LHAVTLNNVAEANFFK" and its annotated spectrum in Figure 1. A fragmentation between "A" and "V" at the $3^{rd}$ and $4^{th}$ positions results in two fragments "LHA-" and "-VTLNNVAEANFFK" that correspond to two ions $b_3$ and $y_{13}$, respectively. Note that b-ions are indexed from the left to the right of the amino acid sequence and y-ions are indexed in the opposite direction. Similarly, a fragmentation between "V" and "T" at the $4^{th}$ and $5^{th}$ positions results in two fragments "LHAV-" and "-TLNNVAEANFFK" and two ions $b_4$ and $y_{12}$. More importantly, the mass difference between $b_3$ and $b_4$ as well as the mass difference between $y_{12}$ and $y_{13}$ match the theoretical molecule mass of the amino acid "V". Hence, given an unannotated spectrum, one could try to iteratively annotate the b-ion and/or y-ion series and predict the corresponding amino acid sequence. The total mass of predicted amino acids should also match the given peptide mass.

However, there are multiple computational challenges. While b-ion and y-ion series are often observed with strong signals, peptide fragmentation actually may produce several types of ions including a, b, c, x, y, z, internal cleavage and immonium ions, as well as their sub-types [3]. Different types of ions have their own intensity distributions and they always mix up together in both directions with respect to the peptide sequence (Figure 1). There are plenty of noise in addition to the real signals, while at the same time, there are also real ions missing from the spectrum, especially for fragmentation near the two terminals of the peptide sequence. Furthermore, we need to take into account measurement errors of mass spectrometers for both peptide and its fragment ions. Those challenges exponentially explode the search space, making it infeasible to identify the right peptide for a spectrum.

There are two key techniques to tackle the problem of peptide identification. If the species information of the biological sample is given, e.g. human, one could search the spectrum against a specific protein sequence database of that species. In this database search problem, the search space is first reduced to a finite set of candidate sequences by filtering based on species and peptide mass. Each candidate is then compared to the spectrum based on a match-scoring function to select the optimum sequence that best fits the fragment ions in the spectrum [6]. If such database information is not available, this becomes a *de novo* sequencing problem, which is much more complicated. Dynamic programming algorithms are often required to explore the huge search space to find the sequence that maximizes their scoring functions [7]. In general, the two techniques, *de novo* sequencing and database search, share the same idea of using some scoring functions to match a spectrum against a peptide, and differ by their search strategies. Another useful information for both techniques is the digestion enzymes because they often cleave proteins with high specificity. For instance, trypsin cuts the peptide bond to the C-terminal side of lysine ("K") and arginine ("R"), and hence, most peptides from trypsin digestion have "K" or "R" at the end of their sequence (Figure 1).

Once the peptides have been identified, it is essential to assemble them back to original protein sequences. This problem is similar to genome or transcriptome assembly from high-throughput sequencing short reads [8]. While protein sequences are much shorter than genome sequences, there are still many obstacles including low coverage, sequencing errors and ambiguities, and the complexity of protein mixture in the biological sample. Assembly algorithms based on overlapping sequences and confidence scores of predicted peptides are the keys to solve this problem [9].

Our study offers a novel deep learning-based method for peptide identification and presents a paradigm shift from algorithm-centric to data-centric approach that greatly benefits from the increasing massive amount of data in proteomics. In the next section, we describe the details of DeepNovo algorithms for protein identification.

## METHODS

### DeepNovo Scoring Function

The scoring function of DeepNovo was first proposed in [1]. The idea is to sequence the peptide by iteratively predicting one amino acid after another and the prediction of the next amino acid depends on the output of previous steps (see Figure 1 in [1]). Mathematically speaking, we calculate a series of conditional probabilities. For example, the score of the sequence "PEPTIDE" given its spectrum is calculated as in the following equation:

$$\begin{aligned}
Prob(PEPTIDE \mid spectrum) = \ & Prob(P \mid spectrum) \\
& * Prob(E \mid P, spectrum) \\
& * Prob(P \mid PE, spectrum) \\
& * Prob(T \mid PEP, spectrum) \\
& * Prob(I \mid PEPT, spectrum) \\
& * Prob(D \mid PEPTI, spectrum) \\
& * Prob(E \mid PEPTID, spectrum) \quad \text{Equation (1)}
\end{aligned}$$

In the implementation, we actually use the log of probability as the score, and the score of a sequence is the sum of its amino acids' scores normalized by its length. Each conditional probability is computed based on two classification models that use the previous output as a prefix to predict the next amino acid. The first model uses a convolutional neural network (CNN) [10-11] to learn features of the intensity distribution of fragment ions in the spectrum. The second model uses a long short-term memory (LSTM) recurrent neural network (RNN) to learn sequence patterns of the peptide [12-15]. The final output of those neural networks is a probability distribution over 20 classes, i.e. amino acid letters, which are used to compute equation (1).

**DeepNovo Database Search**

Knowing the species of the biological sample, we first collect its protein sequences from the UniProt database [16]. We use the taxonomy view to select all strains of the species and download their Swiss-Prot sequences (since those sequences have been carefully annotated and reviewed). Then we perform *in silico* digestion of protein sequences into peptides using the Pyteomics package [17], where the cleavage rules are determined by the enzymes used in the biological experiments. Other options including maximum number of missed cleavages or minimum peptide length are also available. We obtain a list of peptide sequences and compute their theoretical mass which altogether serve as the database for our search.

Given each pair of spectrum and experimental peptide mass from the input data, we need to query them against the database to identify the best-match peptide sequence. The experimental peptide mass is first used to filter the database to a set of candidate sequences with approximately same theoretical mass, subject to a pre-determined error tolerance. The error tolerance is often measured in part-per-million (ppm) and hence is proportional to the experimental peptide mass rather than an absolute amount. Each candidate sequence is then scored against the spectrum using equation (1). It should be noted that while equation (1) shows the calculation in the forward direction, we also implement backward direction and sum up the score of both directions. The benefits of bi-directional sequencing are manifold. First, it reduces the chance of losing information due to missing fragment ions near the two terminals. Second, it allows the LSTM model to learn sequence patterns around the prediction position rather than from one side. Last but not least, using multiple ensembles of the same neural network model often increases its accuracy. The highest-scoring candidate is finally selected as the predicted peptide sequence.

**DeepNovo *De novo* Sequencing**

When database information is not available, we have to perform *de novo* sequencing. This is a global optimization problem where we need to find a peptide sequence such that its total mass is approximately equal to the experimental peptide mass and its matching score against the spectrum is maximized. Unlike database search, it is not possible to reduce the search space to a finite set of candidate sequences. We apply beam search, a heuristic search algorithm that explores a fixed number of top candidate sequences at each iteration. Note that the score based on equation (1) is now computed on partial sequences during the iterative process, and the final predicted sequence is not guaranteed to be the global optimum. The beam search is also accompanied by an off-line dynamic programming algorithm to keep comparing the mass of predicted sequences and the experimental peptide mass during the iterative search. This procedure also helps to filter out those amino acids that do not fit the experimental peptide mass. We also apply bi-directional sequencing here, but in this case, the forward pass and the backward pass may not necessarily produce the same sequence. Overall, *de novo* sequencing is more heuristic and hence less accurate then database search, but this technique can identify novel sequences that do not exist in databases.

**DeepNovo Hybrid Solution**

Since we formulate both techniques, database search and *de novo* sequencing, in the same iterative sequencing approach and use the same scoring function, it is possible to merge them into a hybrid solution. The idea is that if a peptide is novel and does not exist in databases, a *de novo* sequence is likely to have better score than the highest-scoring candidate from database search. This is often not the case for existing methods due to inconsistencies in handling those two problems (e.g. database search tools often incorporate more features into their scoring functions than *de novo* sequencing tools).

## RESULTS

**Performance Evaluation**

We evaluated the accuracy of DeepNovo *de novo* sequencing and database search on a dataset *of Saccharomyces cerevisiae* proteome [18]. The data was acquired from Thermo Scientific Orbitrap Fusion instrument with the Higher-energy Collisional Dissociation (HCD) technique. We first used PEAKS DB software (version 8.0, [4]) to search against the UniProt database with the yeast taxonomy (7,904 protein sequences in total). Other settings include Trypsin digestion, maximum two missed cleavages and one non-specific cleavage, precursor mass tolerance 20 ppm, fragment ion mass tolerance 0.5 Dalton, fixed modification Carbamidomethylation (C), variable modifications Oxidation (M) and Deamidation (NQ), and false discovery rate (FDR) of 1%. The peptide sequences identified from PEAKS DB search were assigned to the corresponding MS/MS spectra and were then used as ground-truth for training and testing the accuracy of DeepNovo. There were 347,047 peptide-spectrum matches in total, 90% of which were used for training, 5% for validation, and 5% for testing. We calculated the total recall as the ratio of the total number of correctly predicted amino acids over the total length of target peptide sequences. We also calculated the recall at the peptide level, i.e. the fraction of target peptide sequences that were fully correctly predicted. Due to the peptide mass constraint, predicted peptide sequences tend to have similar lengths to the corresponding targets, so precision is often very close to recall and we did not report it here.

Table 1 show the *de novo* sequencing and database search results of DeepNovo. We also included the *de novo* sequencing results of PEAKS. It should be noted that PEAKS database search results were used as ground-truth, so the performance of DeepNovo would not exceed that of PEAKS. A more accurate comparison should include several database search tools and use their consensus as ground-truth. The preliminary results here show that DeepNovo identified 89.8% of peptides from PEAKS database search. We further looked into details of the other 10.2% where DeepNovo and PEAKS disagreed. We found multiple evidences that DeepNovo predicted high-quality alternative peptides to complement PEAKS results. Three examples are shown in Figure 2. Interestingly, in the last two examples, DeepNovo predicted the same sequences as PEAKS *de novo* sequencing, while PEAKS database search results were different.

## Table and Figure Legends

Table 1. The recall of PEAKS and DeepNovo at the amino acid level and the peptide level.

|                   | PEAKS *de novo* | DeepNovo *de novo* | DeepNovo database |
|-------------------|-----------------|--------------------|-------------------|
| Amino acid recall | 57.4%           | 74.3%              | 89.9%             |
| Peptide recall    | 25.7%           | 61.7%              | 89.8%             |

Figure 1. An example of peptide, spectrum and annotated fragment ions.

Figure 2. DeepNovo predicts high-quality alternative peptides to PEAKS results.

Figure 1

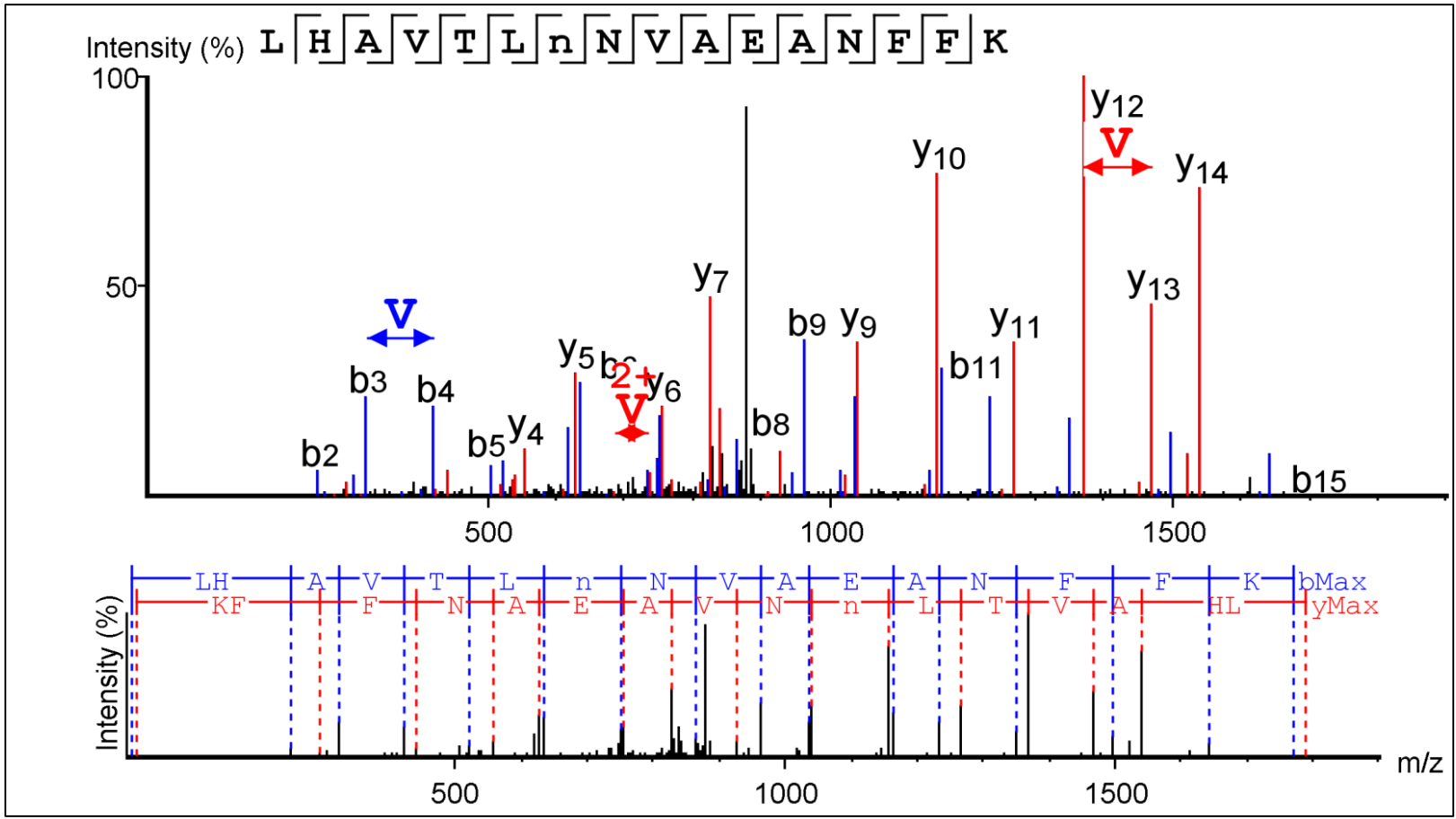

Figure 2a

PEAKS database search

DeepNovo database search

Figure 2b

PEAKS database search

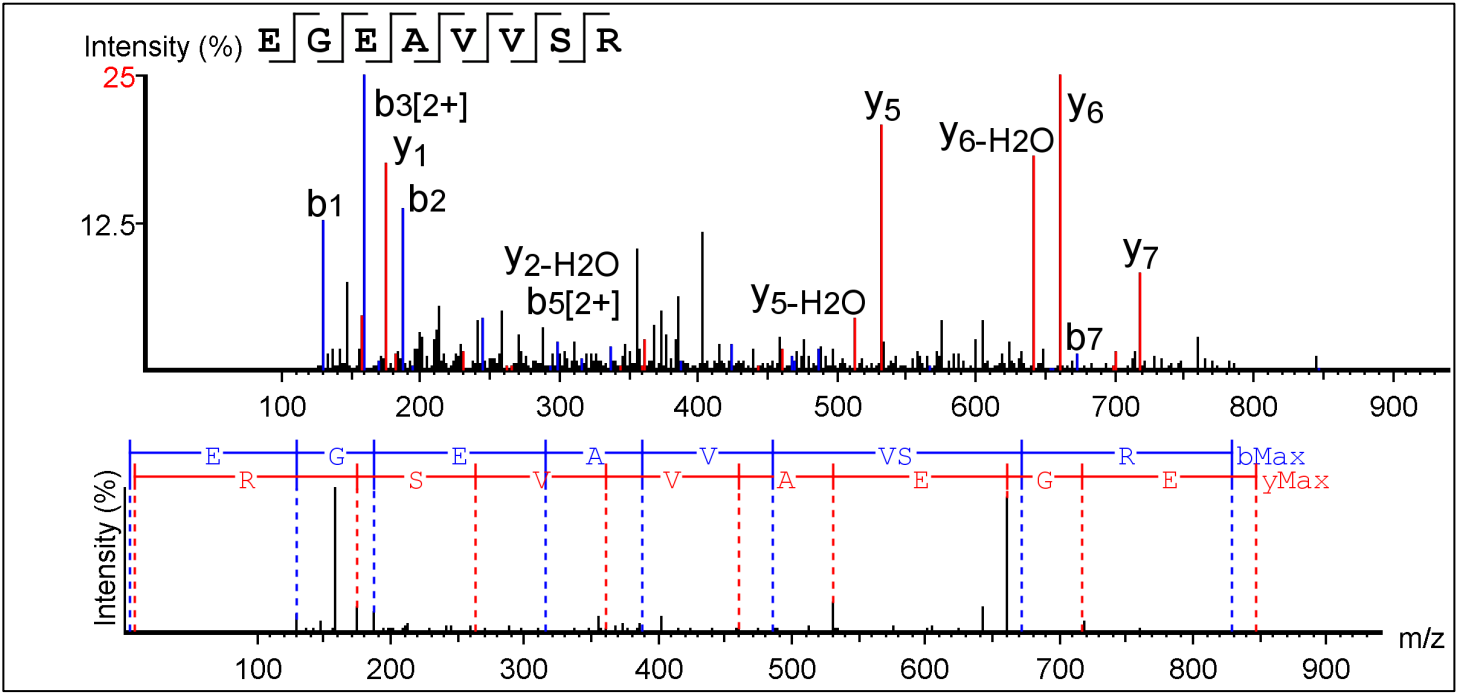

DeepNovo database search

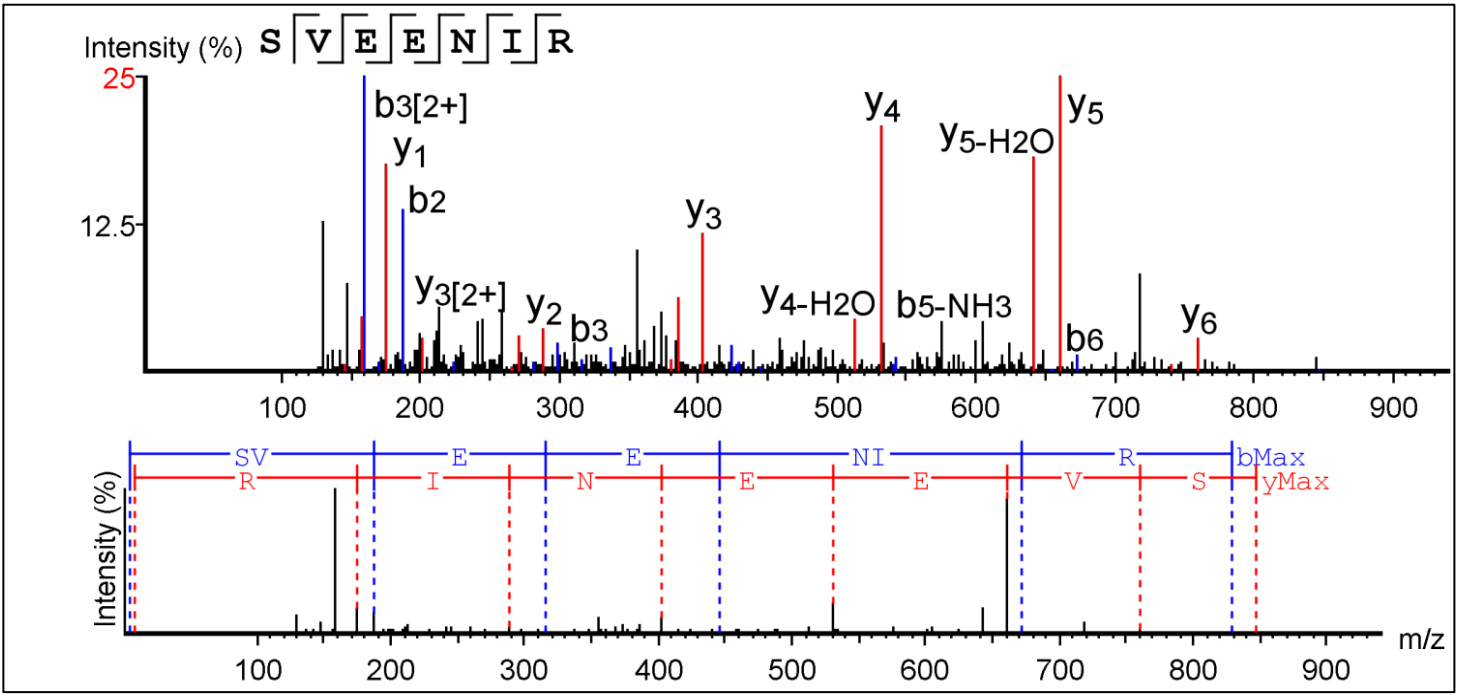

Figure 2c

PEAKS database search

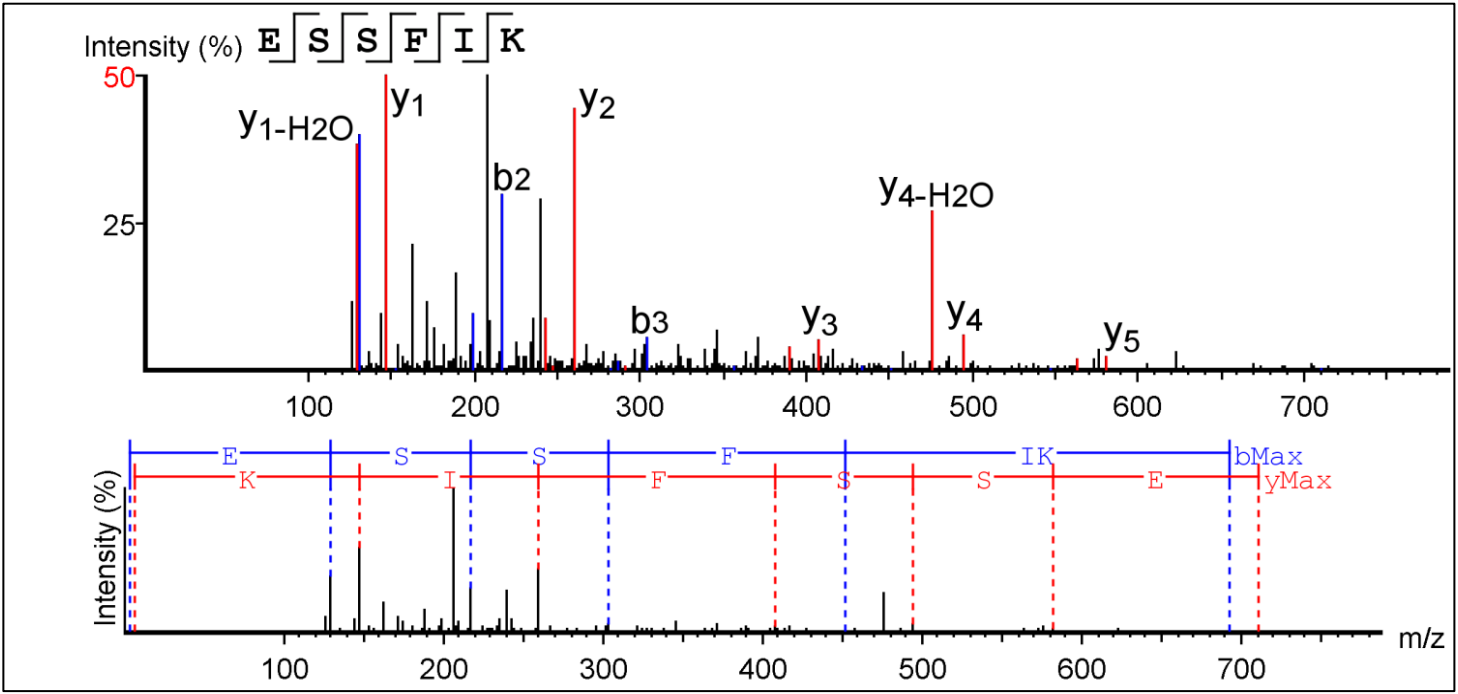

DeepNovo database search

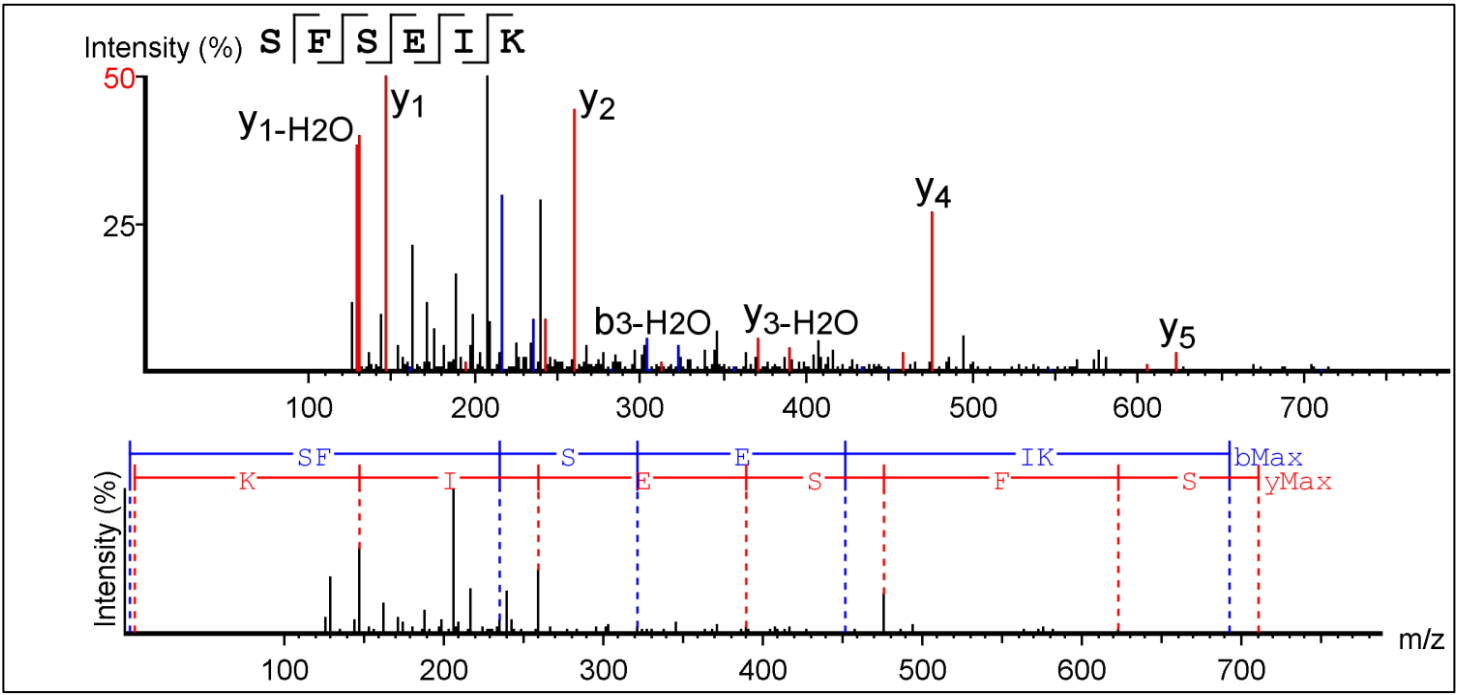